\documentclass[twocolumn,showpacs,nofootinbib,amsmath,amssymb,aps,prd,superscriptaddress,balancelastpage,natbib]{revtex4-1}

\usepackage{graphicx}
\usepackage[caption=false]{subfig}

\usepackage{amsmath,amssymb}
\usepackage{amsfonts,amssymb,mathrsfs}

\usepackage[bookmarks=false,pdfstartview=FitH]{hyperref}
\usepackage[all]{hypcap}

\def\be{\begin{equation}}
\def\ee{\end{equation}}
\def\nn{\nonumber}
\def\f{\frac}
\def\tf{\tfrac}

\def\pl{{\rm Pl}}
\def\lp{\ell_\pl}

\def\d{\dot}

\def\t{\tilde}

\def\dd{{\rm d}}

\def\de{\delta}
\def\ep{\epsilon}

\def\om{\omega}

\def\De{\Delta}
\def\La{\Lambda}
\def\mR{\mathcal{R}}
\def\mH{\mathcal{H}}
\def\mO{\mathcal{O}}

\begin{document}

\pagestyle{plain}

\title{A $\La$CDM Bounce Scenario}

\author{Yi-Fu Cai} \email{yifucai@physics.mcgill.ca}
\affiliation{Department of Physics, McGill University, Montr\'eal, QC H3A 2T8, Canada}

\author{Edward Wilson-Ewing} \email{wilson-ewing@aei.mpg.de}
\affiliation{Department of Physics and Astronomy, Louisiana State University, Baton Rouge, 70803 USA}
\affiliation{Max Planck Institute for Gravitational Physics (Albert Einstein Institute), Am M\"uhlenberg 1, 14476 Golm, Germany, EU}

\begin{abstract}

We study a contracting universe composed of cold dark matter and radiation, and with a positive cosmological constant.  As is well known from standard cosmological perturbation theory, under the assumption of initial quantum vacuum fluctuations the Fourier modes of the comoving curvature perturbation that exit the (sound) Hubble radius in such a contracting universe at a time of matter-domination will be nearly scale-invariant.  Furthermore, the modes that exit the (sound) Hubble radius when the effective equation of state is slightly negative due to the cosmological constant will have a slight red tilt, in agreement with observations.  We assume that loop quantum cosmology captures the correct high-curvature dynamics of the space-time, and this ensures that the big-bang singularity is resolved and is replaced by a bounce.  We calculate the evolution of the perturbations through the bounce and find that they remain nearly scale-invariant.  We also show that the amplitude of the scalar perturbations in this cosmology depends on a combination of the sound speed of cold dark matter, the Hubble rate in the contracting branch at the time of equality of the energy densities of cold dark matter and radiation, and the curvature scale that the loop quantum cosmology bounce occurs at.  Importantly, as this scenario predicts a positive running of the scalar index, observations can potentially differentiate between it and inflationary models.  Finally, for a small sound speed of cold dark matter, this scenario predicts a small tensor-to-scalar ratio.

\end{abstract}


\maketitle

\section{Introduction}
\label{s.intro}

Observations of the cosmic microwave background (CMB) ---most recently \cite{Hinshaw:2012aka, Ade:2013zuv}--- have clearly established that scalar perturbations in the early universe were nearly scale-invariant.  It is thus necessary for any realistic cosmological model to generate, in some fashion, scale-invariant perturbations.

To achieve this, many cosmological models rely on the presence of matter fields (typically scalar fields) that have not yet been observed in nature.  The new matter fields are necessary in these models as they play an essential role in the generation of scale-invariant perturbations.  While it is of course a requirement for any cosmological scenario to predict near scale-invariance in order to be potentially viable, there are some cosmological scenarios where it is possible to avoid the weakness of postulating the existence of unknown matter fields and nonetheless obtain scale-invariance.

We shall study one such model in this paper.  This cosmological model consists of a spatially flat Friedmann-Lema\^itre-Robertson-Walker (FLRW) universe, with a positive cosmological constant, cold dark matter (CDM), and radiation.  These are three ingredients known to be present in our universe, and we will not assume the existence of any other matter fields.  We also assume that the initial conditions are such that the space-time curvature is small and the universe is large and contracting.

As the universe contracts, the space-time curvature will increase, and quantum gravity effects are expected to become important at some point, likely when the space-time curvature nears the Planck scale.  In this work, we will assume that loop quantum cosmology (LQC) captures the salient non-perturbative quantum gravity effects in the very early universe.  LQC, a mini-superspace approach to quantum cosmology motivated by loop quantum gravity, predicts that a bounce occurs near the Planck scale and that, once these quantum gravity effects are included, the space-time is free of the singularities that appear in classical general relativity \cite{Bojowald:2008zzb, Ashtekar:2011ni, Banerjee:2011qu}.

Thus, this model will be that of a bouncing universe, with a matter content of radiation and cold dark matter and a positive cosmological constant.  Now, it is well known in cosmological perturbation theory that, for perturbations that are initially in the quantum vacuum state, the Fourier modes that reach the long wavelength limit in a contracting space-time whose dynamics are dominated by a pressureless matter field become scale-invariant \cite{Wands:1998yp, Finelli:2001sr}.  (Note that the long-wavelength limit of a Fourier mode does not always coincide with the mode exiting the Hubble radius --- here the relevant length scale is the sound Hubble radius, as explained in more detail later.)  Furthermore, if the pressure is slightly negative, for example due to the presence of a positive cosmological constant, then the long wavelength perturbation modes will be almost scale-invariant with a slight red tilt.  Therefore, in the model considered here, we expect the modes that become large during the epoch of the universe that is dominated by cold dark matter to be almost scale-invariant, and those that become large when the effective equation of state is slightly negative to have a small red tilt.

Various realizations of the matter bounce scenario suggested in \cite{Wands:1998yp, Finelli:2001sr} have been considered in the literature, but in most studies the matter content is taken to be scalar fields with specific potentials that can mimic pressureless matter fields for some specific initial conditions.  To the best of our knowledge, this is the first study of a matter bounce scenario where the pressureless matter field is taken to be cold dark matter and that the effect of a positive cosmological constant is also included.  We will compare the predictions of this scenario with other realizations of the matter bounce scenario in Sec.~\ref{s.disc}.

In this paper we calculate the spectrum of the cosmological perturbations for this $\La$CDM bounce scenario.  By the use of some approximations, it is possible to complete the calculations entirely analytically; we also solve the equations numerically in order to provide a check on the validity of the approximations.  In Sec.~\ref{s.hom} we study the dynamics of the background, first analytically and then numerically.  Then in Sec.~\ref{s.scal} we calculate the evolution of the scalar perturbations, from their initial quantum vacuum state to their final form after the bounce in the background FLRW space-time.  Once again, this calculation is first done analytically with the help of some approximations, and is solved numerically afterwards.  We continue in Sec.~\ref{s.tens} by determining the spectrum of the primordial gravitational waves, and end in Sec.~\ref{s.disc} with a discussion.  We use units where $c=1$, but keep $G$ and $\hbar$ explicit except where stated otherwise, typically in the sections devoted to the numerical studies.

\section{Homogeneous Background}
\label{s.hom}

As explained in the Introduction, we are interested in studying the dynamics of a flat FLRW cosmology with a positive cosmological constant $\Lambda$ and whose matter content is composed of radiation and cold dark matter, which are modeled as perfect fluids.

Classically, the dynamics are given by the Friedmann equations, while in LQC there exists a Hamiltonian constraint operator that generates the evolution of the wave function representing the quantum cosmology state.  While the full quantum evolution is in general rather complicated, for sharply peaked states (i.e., states that admit a clear semi-classical interpretation at low curvature scales) the full quantum dynamics are very well approximated by a set of effective equations \cite{Taveras:2008ke, Rovelli:2013zaa, psvt}.  The key point is that since the state is sharply peaked (and it remains sharply peaked throughout the entire evolution, including at the bounce point), it is meaningful to speak of an effective geometry, with an effective scale factor, and to ask what equations of motion govern the dynamics of this effective scale factor; these equations are called the effective equations.  As radiation will dominate the dynamics in the high-curvature regime, it is enough to consider the effective equations for a radiation-dominated flat FLRW space-time, which are given by
\be \label{lqc-fr}
H^2 = \f{8 \pi G}{3} \rho \, \left(1 - \f{\rho}{\rho_c}\right),
\ee
where $H = \d{a}/a$ is the Hubble rate (in proper time), $\rho$ is the energy density of the radiation matter field, and $\rho_c \sim \rho_{\rm Pl}$ is the critical energy density, which is of the order of the Planck energy density.  In addition, the matter field satisfied the continuity equation
\be \label{rad-cont}
\d{\rho} + 4 H \rho = 0,
\ee
where we have used the fact that the pressure of a radiation perfect fluid is $P = \rho/3$.  Note that the classical Friedmann equations are obtained in the limit $\rho_c \to \infty$.  In this paper, we will restrict our analysis to the effective equations of LQC, but a full quantum treatment of a radiation-dominated space-time in LQC is given in \cite{Pawlowski:2014fba}.

In the first part of this section, we will use some reasonable approximations in order to derive some analytical results, and in the second section we present some numerical results that in part complement the analytic results and in part provide a check on the validity of the approximations.

To be specific, we shall make three approximations in the analytical treatment of the background: first, we shall assume that quantum gravity effects are negligible already a few orders of magnitude away from the LQC bounce.  This has been verified in numerical simulations \cite{Pawlowski:2014fba}, and can also be seen from studying the effective equations.  This will allow us to solve the classical Friedmann equations away from the bounce, and the LQC corrections will only become relevant when the space-time curvature nears the Planck scale during the radiation-dominated epoch.

Second, we will assume that the evolution of the background universe can be broken into two distinct eras: the first one which is dominated by the combination of the cosmological constant and cold dark matter, and another era which is dominated by radiation.  We assume a discontinuous change in the equation of state between these two eras, and impose continuity in the scale factor and in the (conformal) Hubble rate during this transition.  This approximation is supported by results in Sec.~\ref{ss.hom-n} that show that the transition between the matter- and radiation-dominated epochs occurs very rapidly.

Finally, recall that the goal of this paper is to calculate the power spectrum of the perturbations.  As is well known, especially from calculations in inflation, the key ingredient that determines the scale-dependence of the perturbations is the equation of state of the background \emph{at the time that the mode reaches the long wavelength limit}.  Therefore, in order to simplify calculations, we will assume a constant equation of state during the time-frame that the perturbation modes of interest reach the long wavelength limit (i.e., when the dynamics of the space-time are dominated by the CDM, but the cosmological constant provides a small correction to the effective equation of state).  This approximation is justified by the effective equation of state being nearly constant during the period of interest, as seen in Sec.~\ref{ss.hom-n}.  Nonetheless, one should keep in mind that the effective equation of state ---due to the combination of $\Lambda$ and CDM--- is in fact changing in time, and in general will be slightly different for different modes.  As we shall see later, this effect leads to a running of the scalar index $n_s$.

\subsection{Analytic Treatment}
\label{ss.hom-a}

Using the effective equations, we shall first determine the dynamics around the bounce point, and then solve for the scale factor at earlier pre-bounce times when the space-time curvature is much smaller.

\subsubsection{Radiation-Dominated Epoch}
\label{sss.rad}

It is easy to see that \eqref{rad-cont} implies that
\be\label{rho_r}
\rho(t) = \f{\rho_o}{a(t)^4},
\ee
where $\rho_o$ is a constant of integration, and this can be used to solve \eqref{lqc-fr}, giving
\be
a(t) = \left( \f{32 \pi G \rho_o}{3} (t-t_o)^2 + \f{\rho_o}{\rho_c} \right)^{1/4}, \nn
\ee
where $t_o$ is another constant of integration.  It is of course possible to choose any values for $t_o$ and $\rho_o$; for convenience we shall set $t_o = 0$ so that the bounce occurs at $t=0$, and $\rho_o = \rho_c$ so that the value of the scale factor at the bounce point is 1.  Then,
\be \label{a-lqc}
a(t) = \left( \f{32 \pi G \rho_c}{3} t^2 + 1 \right)^{1/4}.
\ee

Well before and after the bounce $(|t| \gg \sqrt{3/32 \pi G \rho_c})$, the space-time curvature is much smaller than the Planck scale and the scale factor is very well-approximated by the classical solution
\be \label{a-rad-t}
a(t) = a_o^{1/4} \sqrt{|t|},
\ee
where we have defined
\be \label{a0}
a_o = \f{32 \pi G \rho_c}{3}
\ee
for later convenience, and the Hubble rate is given by
\be \label{hubble-t}
H = \f{1}{2 t}.
\ee

It is also easy in the classical regime to change to conformal time $\eta$ via the relation $a \dd \eta = \dd t$, which gives
\be
|t| = \sqrt \f{2 \pi G \rho_c}{3} \, \eta^2,
\ee
which in turn shows that the scale factor in terms of conformal time is given by
\be \label{a-rad}
a(\eta) = \sqrt \f{8 \pi G \rho_c}{3} \, |\eta|,
\ee
and the conformal Hubble rate, again in the classical regime, is given by
\be \label{hubbles}
\mH = \f{1}{\eta} = - \left( \f{8 \pi G \rho_c}{3} \right)^{1/4} \sqrt{-H};
\ee
the second equality holds for the contracting epoch of the cosmology where $\mH < 0$ and $H < 0$.

\subsubsection{Cold Dark Matter and $\Lambda$}
\label{sss.cdm}

Now we shall consider the earlier epoch where cold dark matter and the cosmological constant dominate the dynamics (the $\Lambda$CDM era).  In this regime, the classical Friedmann equations in conformal time can be written as
\be \label{eq_FRW1}
\mH^2 = \f{8 \pi G}{3} \, a^2 \left( \rho_{CDM} + \rho_{\Lambda} \right) = \f{8 \pi G}{3} \, a^2 \, \rho_{tot},
\ee
where $\rho_{\Lambda} = \Lambda / 8 \pi G$, and the combined continuity equation (also in conformal time) for cold dark matter and the cosmological constant is given by
\be \label{eq_continuity}
\rho'_{tot} + 3 \mH \left( \rho_{tot} + P_{tot} \right) = 0.
\ee
Since $P_\Lambda = - \rho_\Lambda$ and $P_{CDM} = 0$, it follows that the effective equation of state for cold dark matter and the cosmological constant combined%
\footnote{In fact, we expect that $P_{CDM} = \ep^2 \rho_{CDM}$ with $0 < \ep \ll 1$, but here we are interested in the situation where the small positive contribution to $\om$ from the cold dark matter and the small negative contribution to $\om$ from the cosmological constant combine to give a slightly negative $\om$.}
is $P_{tot} = \om \rho_{tot}$, with $-1 \le \om \le 0$.

In order to solve these two equations exactly, we shall assume that $\om = -\de$ is a constant, and furthermore, since we are interested in the regime where the dynamics are dominated by the cold dark matter, we also take $\de \ll 1$.  Of course, the effective equation of state does not remain constant in this setting, but recall that we are interested in calculating the power spectrum of cosmological perturbations, and their scale-dependence depends most sensitively on the effective equation of state at the time when they reach the long wavelength limit. Therefore, the calculations where the spectra of the scalar and tensor perturbations are determined are to be understood as being for the modes that reach the long wavelength limit when the effective equation of state is given by $\om = -\de$, and so the specific value of $\de$ will vary from one mode to another.  Note that this variation will be monotonic with $\de$ becoming closer and closer to zero for shorter and shorter wavelengths, or for larger and larger $k$.  The exact rate at which this occurs will depend on the relative contributions of cold dark matter and the cosmological constant to the total matter energy density.

In the approximation that $\de$ is constant, the total energy density behaves as
\be \label{rho_tot_CDM}
\rho_{tot} = \f{\rho_{eff}}{a^{3(1-\de)}},
\ee
where $\rho_{eff}$ is a constant of integration, and the scale factor is given by
\be
a(\eta) = \left[ \sqrt \f{2 \pi G \rho_{eff}}{3} (1 - 3 \de) (\eta - \eta_o) \right]^{2/(1-3\de)},
\ee
where $\eta_o$ is also a constant of integration.  It follows that the conformal Hubble rate is
\be
\mH = \f{2}{(1-3\de)(\eta-\eta_o)}.
\ee

In order to determine the values of $\rho_{eff}$ and $\eta_o$, let us assume that the transition between the radiation-dominated epoch and the $\Lambda$CDM era occurs at the equality conformal time $\eta_e$.  Then, imposing that the scale factor and the conformal Hubble rate be continuous at the transition time, we find that $\rho_{eff} = \rho_c/a_e^{1+3\de}$ and $\eta_o = \eta_e - 2/[(1-3\de)\mH_e]$, where $a_e = a(\eta_e)$ and $\mH_e = \mH(\eta_e)$ respectively.  Then, the scale factor can be rewritten as
\be \label{a-cdm}
a(\eta) = a_e \left( \f{\eta - \eta_o}{\eta_e - \eta_o} \right)^{2/(1-3\de)}.
\ee

\begin{figure*}[th]
\centering
\subfloat[]
{\includegraphics[width=0.35\textwidth]{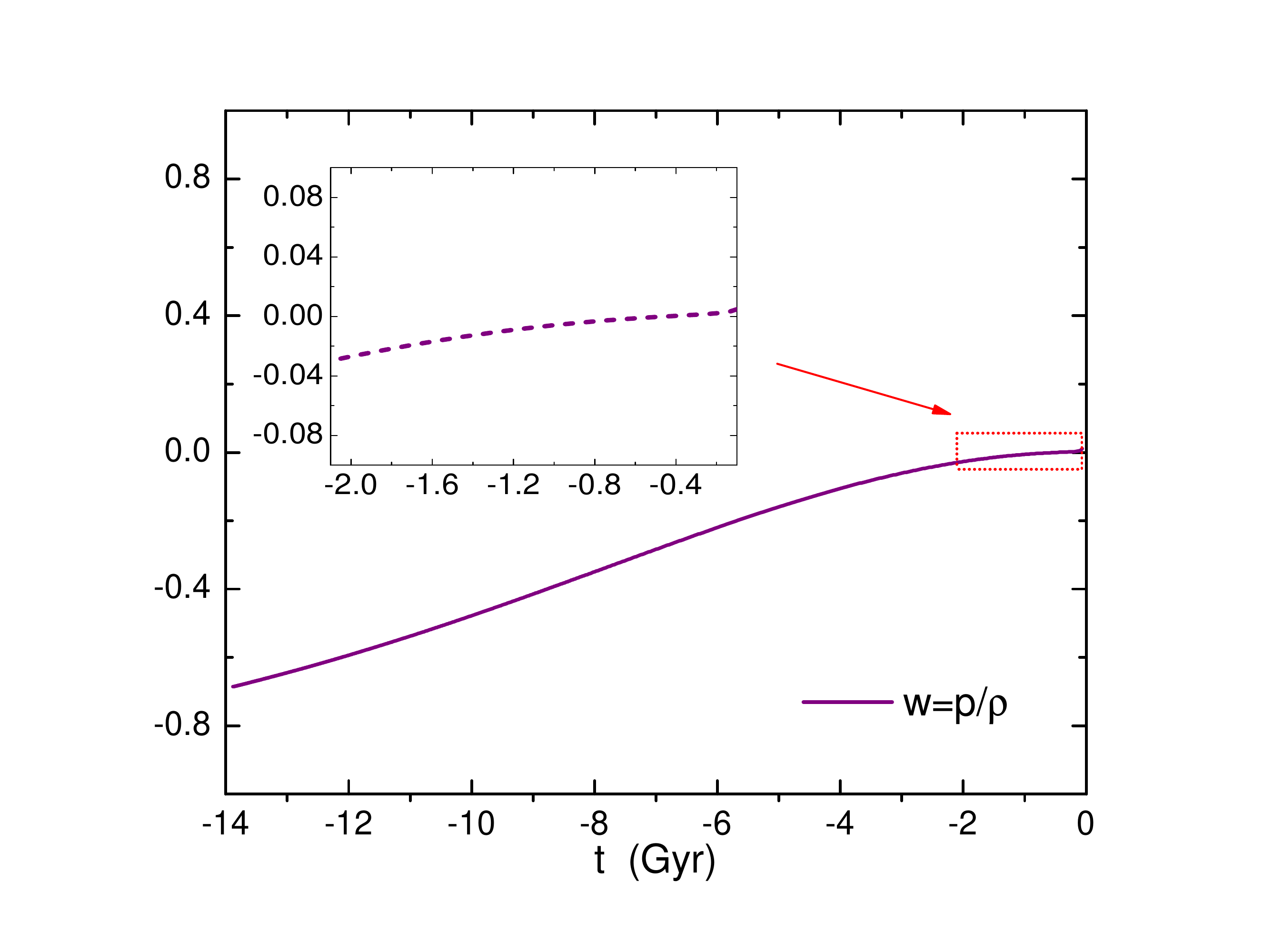} \label{fig1a}}
\subfloat[]
{\includegraphics[width=0.35\textwidth]{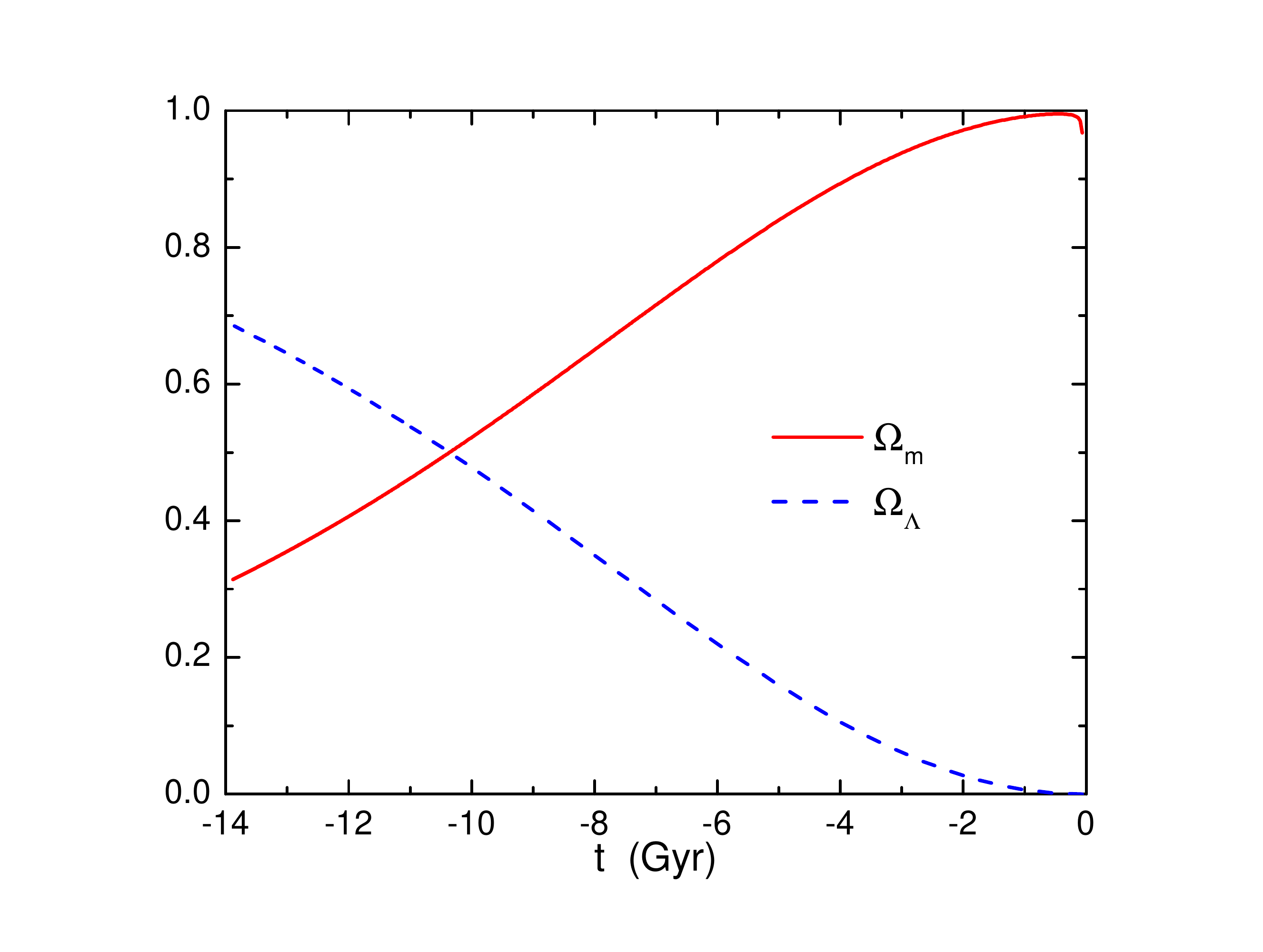} \label{fig1b}}
\caption[]{\footnotesize \hangindent=10pt
Evolutions of the background equation of state parameter $\om$ (the purple solid curve in the left panel), and the density parameters $\Omega_i \equiv \rho_i / \rho_{tot}$ (the red dotted and the blue dashed lines in the right panel) as a function of the cosmic time $t$ (in units of per billion years) in the model under consideration. The horizontal axis denotes the cosmic time $t$. The initial values of background parameters are assumed to be the same as today's universe.
}
\label{fig1}
\end{figure*}

\begin{figure*}[th]
\centering
\subfloat[]
{\includegraphics[width=0.33\textwidth]{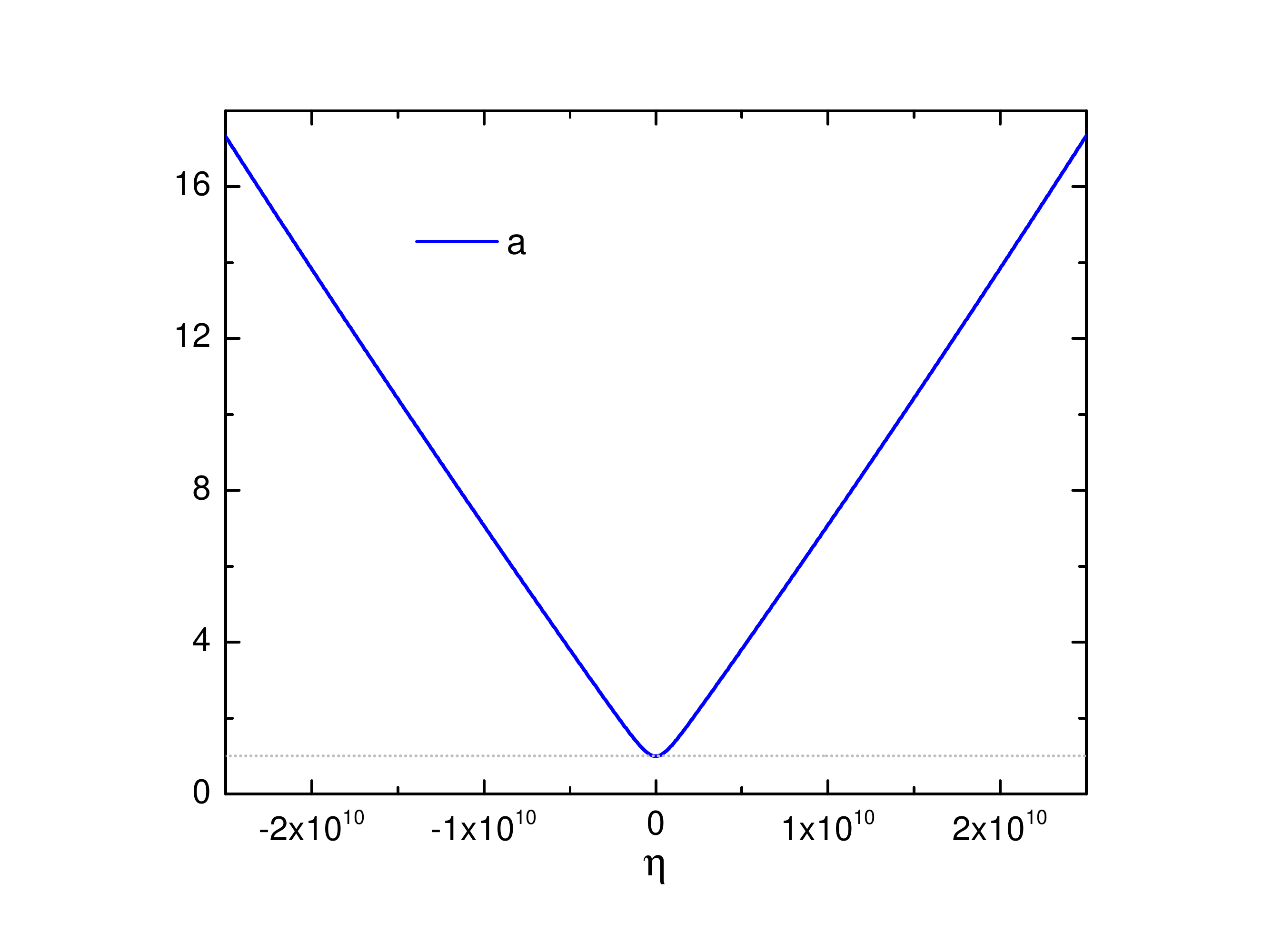} \label{fig2a}}
\subfloat[]
{\includegraphics[width=0.33\textwidth]{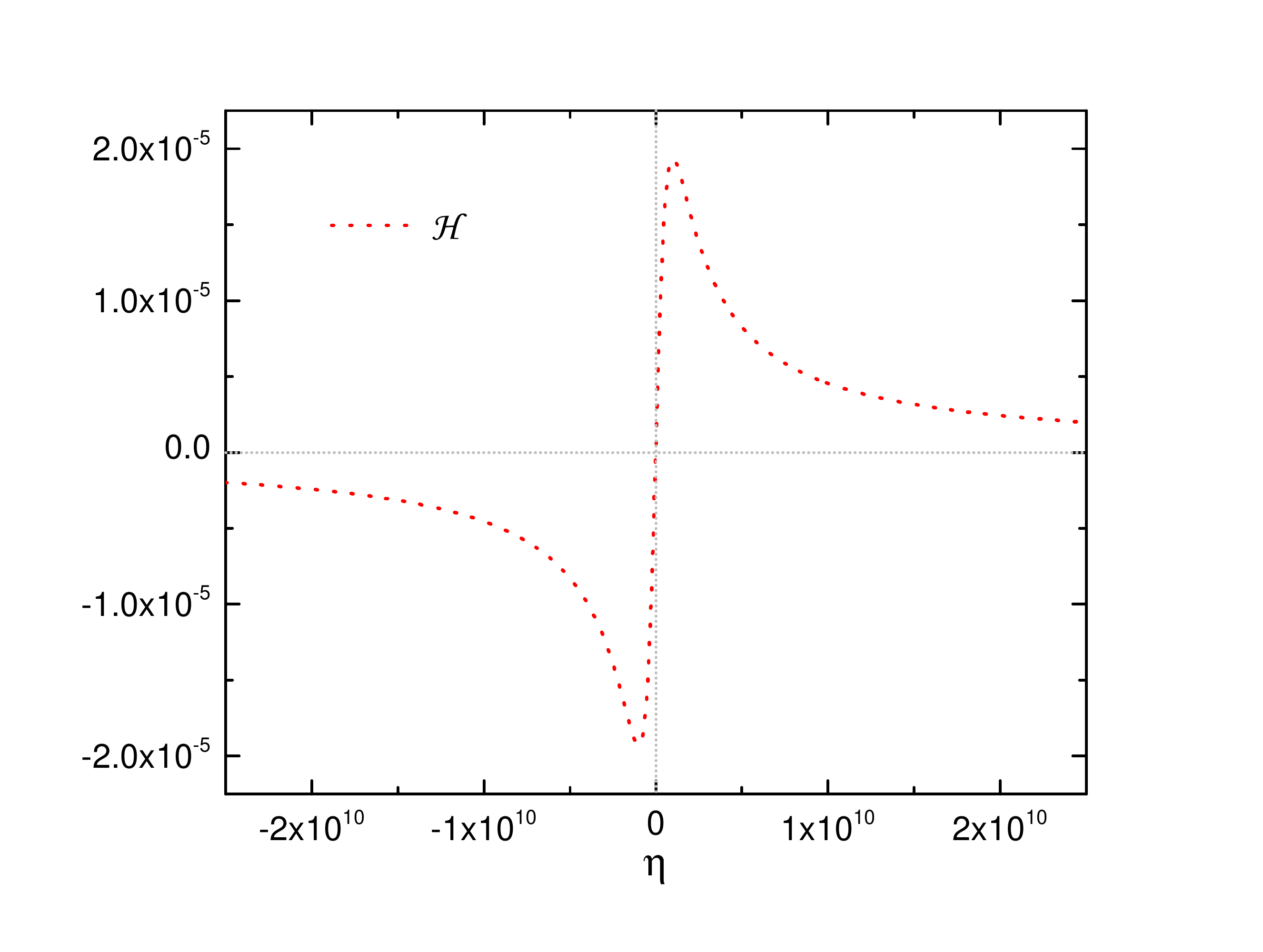} \label{fig2b}}
\subfloat[]
{\includegraphics[width=0.33\textwidth]{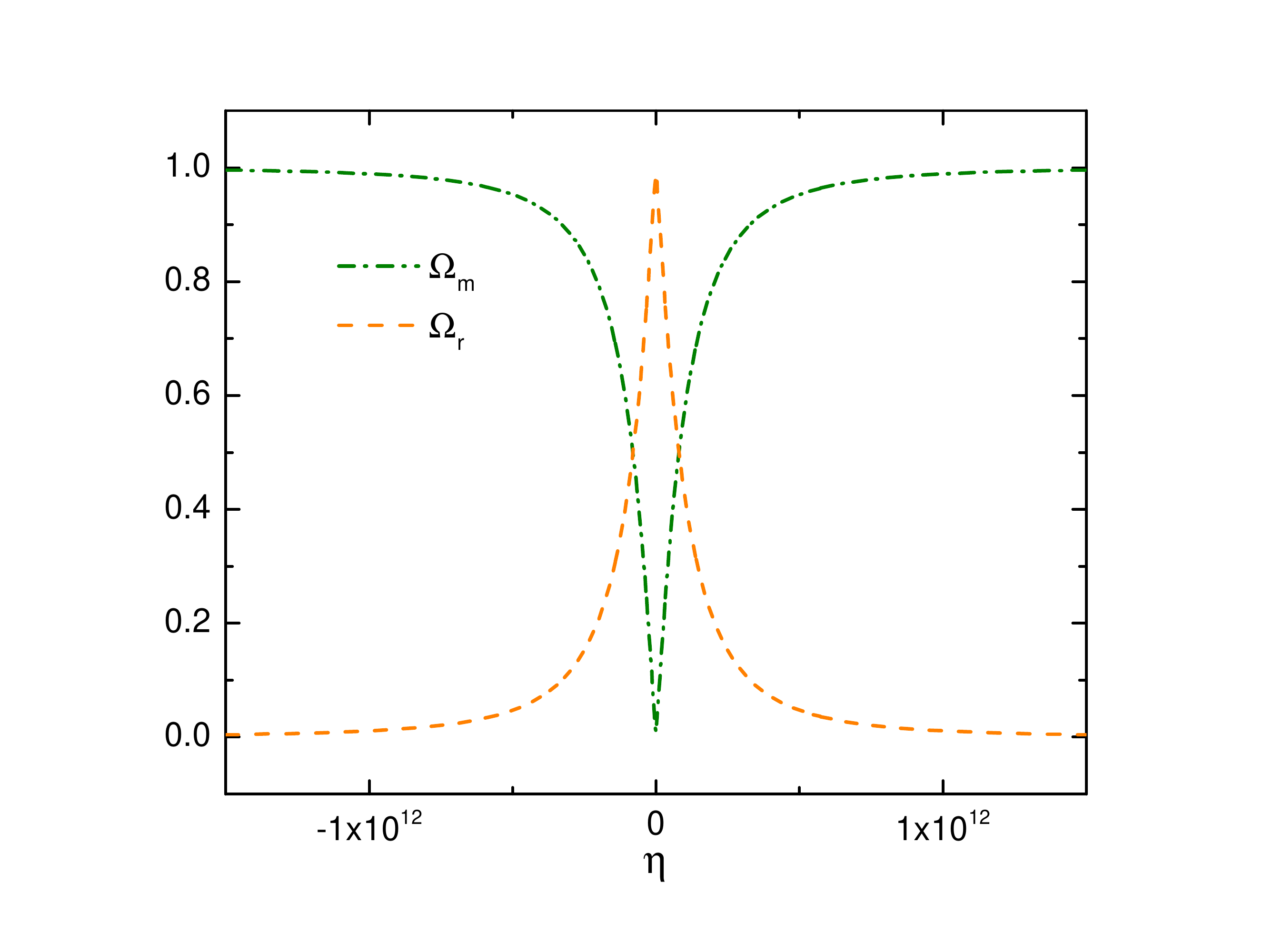} \label{fig2c}}
\caption[]{\footnotesize \hangindent=10pt
Evolutions of the scale factor $a$ (the blue solid curve in the left panel), the conformal Hubble parameter ${\cal H}$ (the red dotted curve in the middle panel) and the density parameters $\Omega_i \equiv \rho_i / \rho_{tot}$ (the green dash-dotted and the orange dashed lines in the right panel) as a function of the conformal time $\eta$ in the model under consideration. The horizontal axis denotes the conformal time $\eta$. The background parameters chosen for the numerics are given in Eq.~\eqref{para_bg}.
}
\label{fig2}
\end{figure*}

\subsubsection{Summary}

Thus, for early times $\eta \le \eta_e$, the scale factor is given by \eqref{a-cdm}; then for $\eta_e \le \eta$ the relation \eqref{a-rad} holds so long as quantum-gravity effects are negligible.  When quantum gravity effects become important, it is necessary to use \eqref{a-lqc} to describe the dynamics of the scale factor.

\subsection{Numerics of the Background Dynamics}
\label{ss.hom-n}

In this subsection we numerically study the background solution presented in the $\Lambda$CDM model within the context of LQC. To be explicit, we study the background universe by separating the evolution into two periods. In the first stage, we apply the realistic data from the Planck results and numerically solve for the period that the contracting universe evolves from dark energy domination to the matter-dominated phase. In particular, we assume that the density parameters of all matter components are the same as what we observed today, which are given by $\Omega_m \equiv \rho_m/\rho_{tot} = 0.314$, $\Omega_\Lambda \equiv \rho_\Lambda/\rho_{tot} = 0.686$ and a deduced value $\Omega_r \equiv \rho_r/\rho_{tot} = 9.23\times 10^{-5}$, respectively \cite{Ade:2013zuv}. During this phase, the numerical result of the background evolution is provided in Fig.\ \ref{fig1}.

One particularly important result is shown in the inset of Fig.\ \ref{fig1a}: the effective equation of state evolves very slowly as the era of matter-domination is approached.  This shows that, for modes that reach the long wavelength limit in this time-frame, any effect due to an evolving effective equation of state will be very small, and it is justified to make the approximation that $\de$ is constant in the analytic calculations.

Then, in the second stage, we consider a toy model of a universe filled with a dark matter component and a radiation component with their energy densities evolving as follows,
\begin{align}
 \rho_m = \rho_m^{i} \left( \frac{a_i}{a} \right)^{3(1-\delta)} ~,~~ \rho_r = \rho_r^{i} \left( \frac{a_i}{a} \right)^4~,
\end{align}
respectively, where $\rho_m^{i}$ and $\rho_r^{i}$ are their initial values when $a=a_i$.  Note that while it would be nice to simultaneously include both matter fields and a cosmological constant, this is significantly more expensive from a computational point of view.  These numerical studies presented here already provide strong support for the approximations used in the analytic section, and we leave more detailed numerical studies for future work.

In order to relate this stage of the dynamics with the previous one, the pressure is taken to have a very small (but non-vanishing) negative value.  Also, to make the comparison between the analytic and numerical results as simple as possible, we adopt the conformal time $\eta$ and set the value of the scale factor $a$ to be unity at the bouncing moment $t_B=0$ in the numerical calculation. The dynamics are calculated by numerically solving the second Friedmann equation ${\cal H}'$ instead of the first one in \eqref{eq_FRW1}, which is given by
\begin{eqnarray}
 {\cal H}' = - \frac{4\pi G}{3} a^2 \left( \rho_{tot} +3 P_{tot} -\frac{4 \rho_{tot}^2}{\rho_c} - \frac{6 \rho_{tot} P_{tot}} {\rho_c}  \right) ~,
\end{eqnarray}
in the case of LQC. In addition, the continuity equation \eqref{eq_continuity} is applied so that the background equations of motion are self-complete after an initial value of the background energy density has been imposed.

For the numerics, we work in units of the reduced Planck mass $M_{\rm Pl} \equiv 1 / \sqrt{8 \pi G}$ (with $\hbar=1$) for all model parameters with dimensions. As an explicit example (although not a realistic one), we choose the values of the energy densities, the critical density $\rho_c$ and an effective equation of state parameter for the CDM at the initial moment to be
\begin{align}\label{para_bg}
 \rho_m^i = 1.1 \times 10^{-24} ~,&~~~
 \rho_r^i = 5.1 \times 10^{-28} ~,~ \nonumber\\
 \rho_c = 2.9 \times 10^{-9} ~,&~~~
 \delta = 0.05 ~,
\end{align}
and our numerical results are shown in Fig.~\ref{fig2}.

The evolution of the scale factor, which is depicted by a blue solid curve in the left panel, explicitly shows that a non-singular bouncing solution is obtained in our model due to the quantum gravity effects captured by LQC, in particular, the minimal value of the scale factor is non-zero.  From the middle panel, one can read more details about the background evolution. For example, the absolute value of ${\cal H}$ is increasing when $\eta$ is about less than $-2\times 10^{9}$. After that, ${\cal H}$ becomes approximately a linear function of the conformal time and correspondingly, the universe enters the bouncing phase with ${\cal H}$ evolving from the negative valued regime to the positive valued one. Eventually, the value of ${\cal H}$ decreases after $\eta \approx 2\times 10^{9}$ and hence the universe naturally connects to a regular thermal expanding phase after the bounce; there is no need for reheating.

The right panel of Fig.~\ref{fig2} characterizes the evolutions of the density parameters of the dark matter and radiation in our model, which are defined by
\begin{align}
 \Omega_{m(r)} \equiv \frac{\rho_{m(r)}}{\rho_{tot}} ~,
\end{align}
with the subscripts ``$_m$" and ``$_r$" representing dark matter and radiation respectively. One can see that the universe was originally dominated by the cold dark matter with $\Omega_m \simeq 1$, as described the green dash-dotted curve.  During the cosmic contraction the contribution of radiation, which is depicted by the orange dashed line, grows faster than that of dark matter and then dominates over the background evolution before the bounce. After the bounce, the universe would have experienced a period of radiation-dominated expanding phase and eventually enters the CDM era and hence is in qualitative agreement with cosmological observations.  Note that the sharp transition between these two eras provides justification for the assumption of a discontinuous transition between matter- and radiation-domination used in the analytic calculations.  Also, it is important to keep in mind that a more careful choice of the initial parameters would make the model more precisely consistent with experimental data.

\section{Scalar Perturbations}
\label{s.scal}

In this section, we will calculate the final spectrum of scalar perturbations after the bounce, assuming they begin in the quantum vacuum state in the distant past of the pre-bounce epoch.

In cosmological perturbation theory, it is convenient to use the gauge-invariant Mukhanov-Sasaki variable \cite{Mukhanov:1990me}
\be
v = z \, \mR,
\ee
where $\mR$ is the comoving curvature perturbation and
\be
z = \f{a \, \sqrt{\rho+P}}{c_s \, H}.
\ee
Linear perturbations can be handled in LQC by following the `separate universe' approach presented in \cite{Salopek:1990jq, Wands:2000dp}, and from the resulting quantum theory it is possible to derive effective equations that can be used to calculate expectation values of sharply-peaked states.  The LQC effective equations for the Mukhanov-Sasaki variable are \cite{WilsonEwing:2011es, WilsonEwing:2012bx}
\be \label{lqc-ms}
v'' - c_s^2 \, \left(1 - \f{2 \rho}{\rho_c} \right) \nabla^2 v - \f{z''}{z} v = 0,
\ee
and it is easy to see that the standard classical expression is recovered in the limit $\rho_c \to \infty$.  The effective equation \eqref{lqc-ms} is expected to provide a good approximation to the full quantum dynamics for modes that always remain large compared to the Planck length \cite{Rovelli:2013zaa}, which is the case for the observationally relevant modes in the matter bounce scenario.

In this section, using \eqref{lqc-ms} we will determine how the Fourier modes $v_k$ evolve, from their initial quantum vacuum state, to their form as they exit the sound Hubble radius%
\footnote{Since the sound speed of the matter fields in this model is not 1, we find that the Fourier modes reach the long wavelength limit when the mode exits the sound Hubble radius, not the Hubble radius.  These quantities differ by a factor of $c_s$.  Note that it is equivalent to state that a given mode reaches the long wavelength limit when its `sound wavelength' exits the Hubble radius.}
$r_{sH} = c_s / H$ and finally how they propagate through the bounce.  As we shall see, the Fourier modes that exit the sound Hubble radius during the period of matter domination in the contracting branch become scale-invariant and therefore these modes are of particular interest.  This is why in this paper we will only consider this family of the Fourier modes $v_k$, and ignore the modes that exit the Hubble radius either before (during the epoch dominated by the cosmological constant) or after (during the radiation-dominated phase).

In the first part, we present analytical calculations, and in the second, numerical simulations.  As in the previous section, it is necessary to make certain approximations in order to make the analytical calculations tractable and the numerical studies in the second part serve in part to check that the approximations are valid.

\subsection{Analytic Treatment}
\label{ss.scal-a}

We will begin by solving the dynamics of the perturbations in the $\Lambda$CDM era ---and the modes of interest are those that reach the long wavelength limit during this era--- and then determine their evolution during the radiation-dominated epoch, including through the bounce.

In order to solve the equations of motion for $v_k$, it is necessary to make the following approximations: (i) we only consider the modes that reach the long wavelength limit during the $\Lambda$CDM era, where the background matter field is modelled as a perfect fluid with a small and negative constant equation of state as explained in Sec.\ \ref{sss.cdm}; (ii) we assume continuity in $v_k$ and $v_k'$ at the spatial slice where we approximate the equation of state changing discontinuously to the radiation-dominated epoch; and (iii) we work in the long wavelength limit during the bounce period.

Recall from Sec.\ \ref{ss.hom-n} that the equation of state changes very slowly in the regime where the effective equation of state is slightly negative.  This supports the first approximation, since the key ingredient in determining the long wavelength spectrum of scalar perturbations is the equation of state of the background at the time that the given mode exits the sound Hubble radius.  Since the effective equation of state is changing slowly, we expect corrections to the approximation of a constant equation of state to be subleading.  Finally, the validity of approximations (ii) and (iii) is verified numerically in Sec.\ \ref{ss.scal-n}.

\subsubsection{The $\Lambda$CDM Era}
\label{sss.perts-cdm}

For the contracting portion of the space-time where the scale factor is given by \eqref{a-cdm} (and safely neglecting quantum gravity effects at this stage), \eqref{lqc-ms} is
\be
v''_k + c_s^2 \, k^2 v_k - \f{2(1+3\de)}{(1-3\de)^2 (\eta-\eta_o)^2} v_k = 0.
\ee
Since the sound speed of cold dark matter is unknown, we set $c_s = \ep$ which we assume to be constant.  We expect $\ep$ to be a small positive number.

The solutions to this differential equation are
\begin{align}
v_k = \sqrt{-(\eta-\eta_o)} & \Big( A_1 H_n^{(1)}[-\ep k (\eta-\eta_o)] \nn \\
& + A_2 H_n^{(2)}[- \ep k (\eta-\eta_o)] \Big),
\end{align}
where $H_n^{(1)}$ and $H_n^{(2)}$ are the Hankel functions, and
\be
n = \sqrt{ \f{2(1+3\de)}{(1-3\de)^2} + \f{1}{4} } \approx \f{3}{2} + 6 \de + O(\de^2),
\ee
where after the last equality we drop terms of order $\de^2$ and higher (recall that $\de \ll 1$).

Choosing the initial conditions to be quantum vacuum fluctuations sets $A_1 = \sqrt{\pi \hbar / 4}$ and $A_2 = 0$.

Then, as $\eta$ approaches $\eta_e$, some modes satisfy $- \ep k (\eta-\eta_o) \ll 1$.  These modes are said to be in the long wavelength limit, and in this limit it is possible to use the small argument expansion of the Hankel functions to show that
\begin{align} \label{v-cdm}
v_k &= \sqrt \f{-\pi \hbar(\eta-\eta_o)}{4} \Bigg[ \f{(\ep k)^n}{\Gamma(n+1)} \left( \f{-(\eta-\eta_o)}{2} \right)^n \nn \\
& \qquad - i \f{\Gamma(n)}{\pi (\ep k)^n} \left( \f{-2}{\eta-\eta_o} \right)^n \Bigg] \\
&= \sqrt \f{8 \hbar}{9} (\ep k)^{3/2 + 6\de} \mH^{-(2+6\de)} \nn \\
& \qquad - i \sqrt \f{\hbar}{4} (\ep k)^{-3/2 - 6\de} \mH^{1+6\de},
\end{align}
where in the second equality the time dependence has been rewritten in terms of the conformal Hubble rate, and the exponents are accurate to first order in $\de$, while the numerical prefactors are only accurate to zeroth order in $\de$.  It is straightforward to determine higher order corrections in $\de$, but this will not be necessary here.

\subsubsection{Radiation-Dominated Epoch}
\label{sss.perts-rad}

During the radiation-dominated epoch, the scale factor is proportional to $\eta$ while the sound speed is given by $1/\sqrt{3}$, and so the Mukhanov-Sasaki variable satisfies the equation
\be
v_k'' + \f{k^2}{3} v_k = 0,
\ee
at least in the classical regime where quantum gravity effects are negligible.  Note that due to the drastic change in the speed of sound, some modes that were in the long wavelength regime may at first be in the short wavelength limit at the onset of the radiation-dominated epoch.

Therefore, the relevant solutions are
\be \label{v-rad1}
v_k = B_1 \sin \f{k \eta}{\sqrt 3} + B_2 \cos \f{k \eta}{\sqrt 3},
\ee
and $B_1$ and $B_2$ can be determined from \eqref{v-cdm} by demanding that $v_k$ and $v'_k$ be continuous at $\eta_e$.  Note that as the bounce is approached $\eta \to 0$ and therefore the second term with the prefactor $B_2$ will dominate, so we can drop $B_1$.  Imposing continuity in $v_k$ and $v'_k$ gives
\begin{align} \label{b2}
B_2  = & - i \sqrt \f{\hbar}{4} (\ep k)^{-3/2-6\de} \cos \f{k \eta_e}{\sqrt 3} \,\, \mH_e^{1+6\de} \nn \\
& - i \sqrt \f{3 \hbar}{16} (\ep k)^{-3/2-6\de} k^{-1} \sin \f{k \eta_e}{\sqrt 3} \,\, \mH_e^{2+6\de}.
\end{align}
Some modes will have rebecome short wavelength modes due to the drastic change in the sound speed.  For these modes, the small argument expansion for the trigonemetric functions cannot be used, and these terms will not be scale-invariant.  Thus, we expect that scale-invariance will only be obtained for the modes that satisfy $k \eta_e \ll 1$.  We will return to this point later.

\subsubsection{The Bounce}
\label{sss.perts-bounce}

In the contracting phase, as the bounce is approached (but before quantum gravity effects become important) we have $| k \eta | \ll 1$ and in this limit the solution for the Mukhanov-Sasaki variable \eqref{v-rad1} tends to
\be \label{v-rad-bef}
v_k = B_2,
\ee
with $B_2$ given in \eqref{b2}.

During the bounce, all of the modes of cosmological interest remain in the long-wavelength limit, and therefore the equation of motion for $v_k$ is
\be
v_k'' - \f{z''}{z} v_k = 0,
\ee
where $z = a \sqrt{\rho + P} / c_s H = 4 \sqrt{\rho_c} \, a^3 / (a_o \, t)$ [recall that $c_s = 1/\sqrt{3}$ during radiation-domination and $a_o$ is defined in \eqref{a0}], and the solution is
\be
v_k = C_1 z + C_2 z \int_\eta \f{\dd \t\eta}{z(\t\eta)^2}.
\ee
Note that $z$ is not simply proportional to $a$ due to the quantum gravity effects that modify the Friedmann equation at high curvatures as seen in \eqref{lqc-fr}.  The integral can be evaluated by rewriting it in terms of the proper time via the relation $\dd t = a \, \dd \eta$, giving
\be \label{v-2f1}
v_k = C_1 \, z + C_2 \, z \, t \left[ {}_2F_1 \left(\tf{1}{2},\tf{3}{4};\tf{3}{2},-a_o t^2 \right) - \f{1}{a^3} \right],
\ee
where $C_2$ has been redefined in order to absorb some numerical factors.

In the classical pre-bounce era ($t \ll - \sqrt{a_o}$), this expression must agree with \eqref{v-rad-bef} and this uniquely determines
\be \label{c_i}
C_1 = \pi \sqrt\f{G}{6} \, \f{\Gamma\left(\tf{1}{4}\right)}{\Gamma\left(\tf{3}{4}\right)} \, B_2, \qquad
C_2 = - \sqrt \f{2 \pi G a_o}{3} \, B_2.
\ee
Note that during this calculation it is important to keep in mind that $t = - |t|$ in the pre-bounce era.

It is also easy to calculate the form of the scalar perturbations in the classical post-bounce era by taking the limit $t \gg \sqrt{a_o}$ in \eqref{v-2f1}, which gives in terms of the comoving curvature perturbation
\be \label{r-fin}
\mR_k = \f{v_k}{z} = 2 C_1 + \mO\left(t^{-1}\right),
\ee
where we have only kept the dominant contribution, namely the constant mode which is the only one that does not decay with time.

\subsubsection{Results}
\label{sss.perts-res}

The amplitude of the comoving curvature perturbation of $2 C_1$ after the bounce depends on $B_2$ via \eqref{c_i} and so it is easy to check whether the resulting scalar perturbations \eqref{r-fin} are scale-invariant or not.  Since the only dependence of $k$ in $C_1$ resides in $B_2$, a quick examination of \eqref{b2} suffices to determine the scale-dependence of $\mR$.  As one can readily verify, one obtains near scale-invariance only in the limit of $|k \eta_e| \ll 1$ (otherwise the dominant contribution would be oscillations superimposed over a red spectrum), in which case
\be
B_2 = -\f{3 \, i}{4} \sqrt\hbar \, (\ep k)^{-\tf{3}{2}-6\de} \mH_e^{1+6\de} \left( 1 + \mO\left(k^2 \eta_e^2\right) \right).
\ee
Therefore, a necessary condition for scale-invariance is that
\be \label{k-lim}
\left| k \eta_e \right| \ll 1,
\ee
that is to say that the modes that become (nearly) scale-invariant during the matter-dominated contracting era must remain outside the sound Hubble radius during the entire contracting radiation-dominated epoch and the bounce in order to remain (nearly) scale-invariant.

For these modes, the power spectrum is
\begin{align} \label{spectrum}
\De_\mR^2 =& \f{k^3}{2 \pi^2} |\mR|^2 \nn \\
=& \sqrt\f{3 \pi}{2} \,
\left( \f{\Gamma\left(\tf{1}{4}\right)}{\Gamma\left(\tf{3}{4}\right)} \right)^2 \,
\sqrt\f{\rho_c}{\rho_{\rm Pl}} \cdot \f{|H_e| \lp}{\ep^3} \nn \\ & \quad \times
\left( \f{8 \pi G \rho_c |H_e|}{3 k^4} \right)^{3 \de},
\end{align}
where $\rho_{\rm Pl} = 1/(G^2\hbar)$ and $\lp = \sqrt{G\hbar}$, and the tilt is given by
\be \label{tilt}
n_s = 1 - 12 \, \de.
\ee
Thus, the observation of the tilt to be $n_s \approx 0.96$ \cite{Hinshaw:2012aka, Ade:2013zuv} sets $\de \approx 0.003$, which means that when the wavelength $k^{-1}$ exits the sound Hubble radius $\ep / H$ the effective equation of state must have been $\om_{\rm eff} \approx -0.003$.

Also, in this model we predict a small running of the scalar index.  This is due to the following two effects: (i) the departure from scale-invariance in a small interval of $k$ depends on the background effective equation of state, and (ii) the background equation of state is dynamical.  The smallest values of $k$ reach the long-wavelength limit first, at a time when $\La$ contributes slightly more to the background dynamics than it does at later times when larger values of $k$ reach the long-wavelength limit.  Therefore, as $k$ increases, the background equation of state at the `sound-Hubble-crossing' time also increases,
\be
\f{\dd \om_{\rm eff}}{\dd k} > 0,
\ee
and since $\om_{\rm eff} = - \de$ and $n_s = 1 - 12 \de$, it follows that
\be \label{running}
\f{\dd n_s}{\dd k} > 0.
\ee
Therefore another prediction of this realization of the matter bounce scenario is for the scalar index $n_s$ to increase with $k$.  Although the presence of this effect is clear in this model, its amplitude is not known.  In order to calculate the expected amplitude of $\dd n_s / \dd k$, it would be necessary to know quite precisely the energy densities corresponding to cold dark matter and the cosmological constant during the contracting phase, which is not an easy task especially since we do not expect the universe to be symmetric around the bounce point, as we shall discuss next.

Nonetheless, the qualitative result \eqref{running} is a clear prediction for this realization of the matter bounce scenario.

Finally, as stated above, the results \eqref{spectrum} and \eqref{tilt} only hold for Fourier modes that satisfy the condition \eqref{k-lim}.  In order to better understand this condition, it is useful to rewrite it in terms of the physical wave number $k = a(t) \cdot k_{\rm phys}(t)$ and of the Hubble rate at equality $H_e$ via \eqref{a-rad-t}, \eqref{hubble-t} and \eqref{hubbles},
\be \label{k-phys}
\f{k_{\rm phys}(t_e)}{|H_e|} \ll 1.
\ee
A good choice to ensure that $k_{\rm phys}$ corresponds to modes that are observed in the cosmic microwave background today, is to choose%
\footnote{Of course, this relation has to hold for all observed $k$, here we simply choose a reasonable value of $k$ in order to better understand the consequences of imposing \eqref{k-phys}.}
$k_\star = 0.05 \: {\rm Mpc}^{-1} = 10^{-59} \lp^{-1}$, which lies roughly in the middle of the logarithmic range of the scales probed by the Planck telescope \cite{Ade:2013zuv}.  The value of $k_\star$ at the time of equality is given by the relation $k_\star(t_e) = a(t_o) \cdot k_\star(t_o) / a(t_e)$.  With our choice of conventions of setting $a(t=0) = 1$ at the bounce, it follows that $a(t_o) \sim 10^{31}$.  This can be calculated from the fact that (i) the scale factor increased by a factor of $\sim 10^4$ after matter-radiation equality until today, and (ii) that in the expanding branch matter-radiation equality is known to occur at $t_e^+ \sim 10^4 \, {\rm years} \sim 6 \times 10^{54} t_{\rm Pl}$ giving a scale factor of $a(t_e^+) = a_o^{1/4} \sqrt{t_e} \sim 10^{27}$ (assuming $\rho_c \sim \rho_{\rm Pl}$) at the time of matter-radiation equality in the expanding branch.  (Recall that matter-radiation equality occurs before recombination, and the superscript `+' on $t_e^+$ denotes the matter-radiation equality in the \emph{expanding} post-bounce branch.)

\begin{figure*}[th]
\centering
\subfloat[]
{\includegraphics[width=0.33\textwidth]{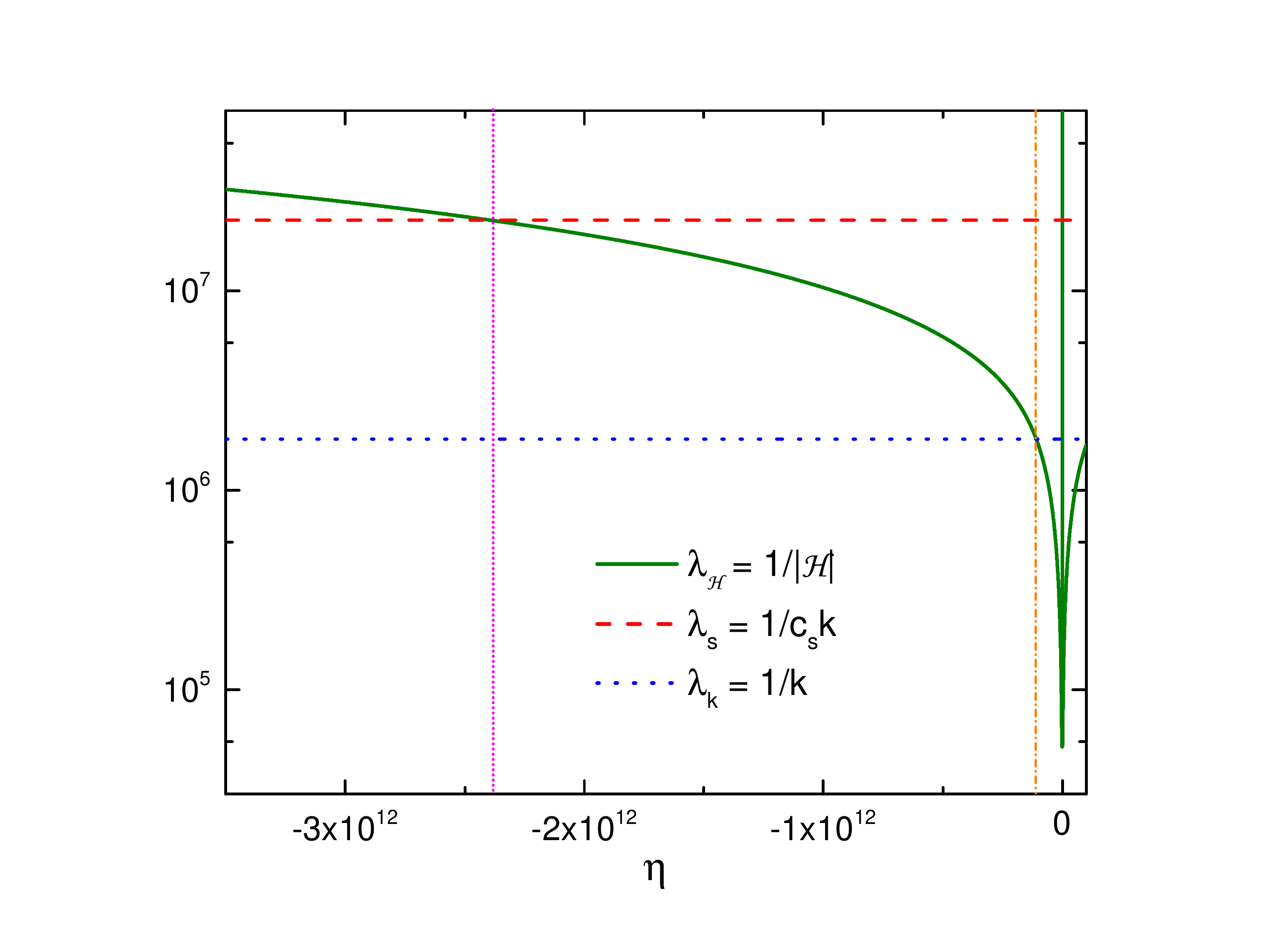} \label{fig3a}}
\subfloat[]
{\includegraphics[width=0.33\textwidth]{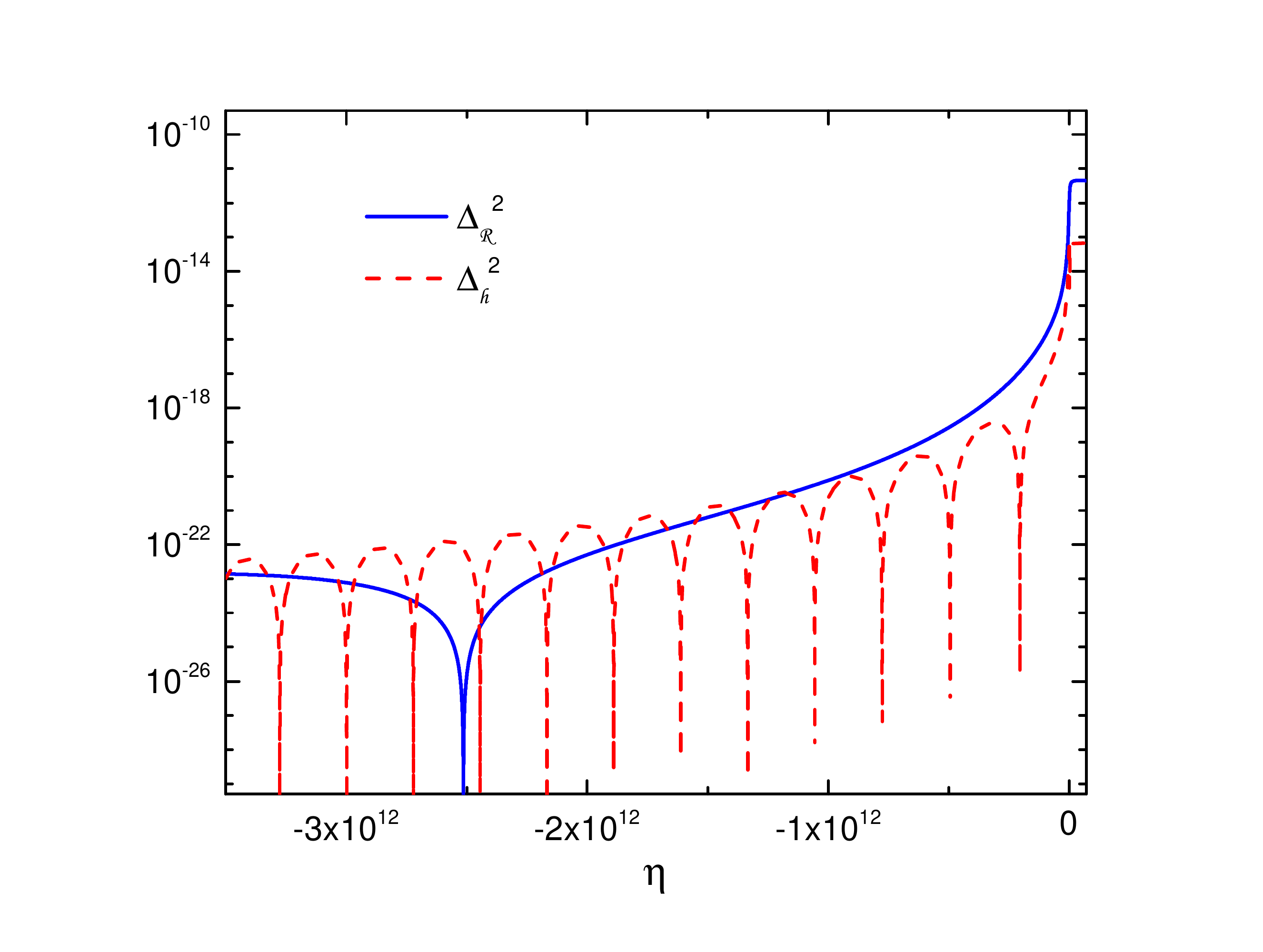} \label{fig3b}}
\subfloat[]
{\includegraphics[width=0.33\textwidth]{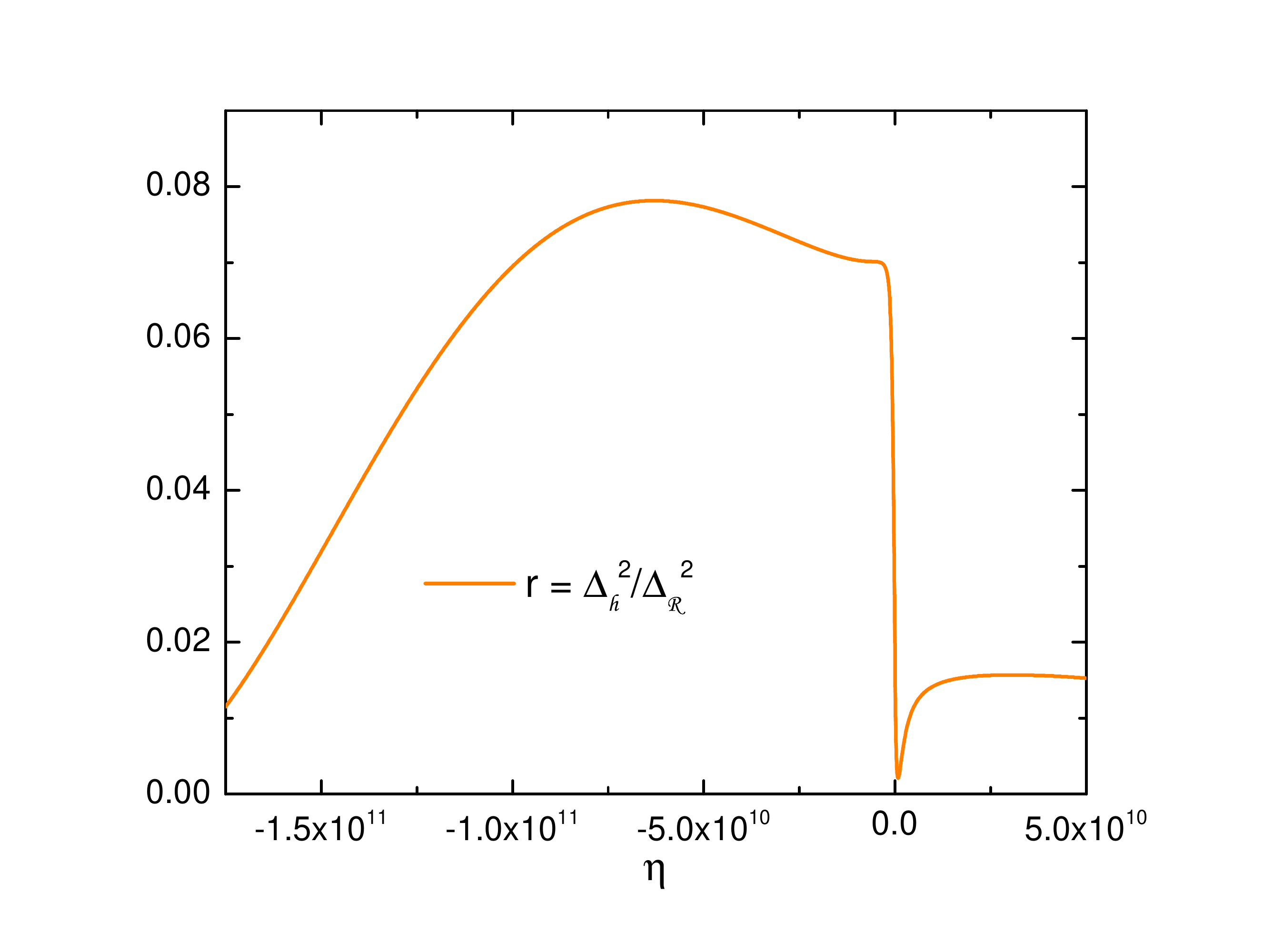} \label{fig3c}}
\caption[]{\footnotesize \hangindent=10pt
Evolutions of cosmological perturbations with a fixed comoving wave number $k=2.2 \times 10^{-7}$ in the model under consideration in the frame of LQC. The left panel shows the comparison among the conformal Hubble radius $\lambda_{\cal H}=1/{\cal H}$ (the green solid line), the comoving sound wavelength $\lambda_{\ep}=1/{\ep k}$ (the red dashed line), and the regular comoving wavelength $\lambda_k=1/k$ (the blue dotted line). The middle panel depicts the dynamics of primordial power spectra for curvature perturbations $\Delta_{\cal R}^2$ (the blue solid curve) and gravitational waves $\Delta_h^2$ (the red dashed curve) along with the bouncing background. The right panel presents the evolution of the tensor-to-scalar ratio $r$ (the orange solid line). The initial conditions for the background field and model parameters are the same as for Fig.~\ref{fig2} and are given in \eqref{para_bg}. The initial conditions for cosmological perturbations of both scalar and tensor types are that they are initially quantum vacuum fluctuations in the early matter-dominated contracting phase.
}
\label{fig3}
\end{figure*}

If we assume a symmetric bounce, then it follows that the times of matter-radiation equality before and after the bounce are symmetric around $t=0$, in which case $t_e = -6 \times 10^{54} t_{\rm Pl}$.  From this and the relation \eqref{hubble-t}, it is easy to check that $k_{\rm phys} / |H_e| \sim 1$ is of the order of unity rather than much smaller than 1.  This shows that a symmetric bounce is not viable in this model.  To make this conclusion explicit, we rewrite \eqref{k-lim} as
\be
\f{a(t_e^+) \, H_e^+}{a(t_e) \, H_e} \cdot \f{a(t_o) \, k_\star(t_o)}{a(t_e^+) \, H_e^+} \ll 1,
\ee
which in turn, since the second term is of order unity and via \eqref{a-rad-t} and \eqref{hubble-t}, gives
\be \label{asymmetry}
\f{a(t_e)}{a(t_e^+)} \ll 1.
\ee
This relation shows that the bounce must be significantly asymmetric.  Indeed, in order for the condition \eqref{k-phys} to hold, $|H_e|$ must be much larger (by at least a few orders of magnitude) than it would be in a model with a symmetric bounce, in which case it is necessary for the matter-radiation equality to occur at much higher curvature scales in the contracting branch than it does in the expanding branch.  One possible way for this to happen would be if a very large number of additional quanta of radiation are created during the bounce; as an aside note that such a process would generate a significant amount of entropy.

Interestingly, it has recently been suggested that particle production may play an important role during the bounce in LQC and this effect would cause the bounce to be significantly asymmetric in precisely the manner outlined here \cite{Mithani:2014jva}.  That being said, it is not yet clear whether particle production could generate the amount of asymmetry that is required for this realization of the matter bounce to be viable.  We leave this question for future work.

There are also other reasons why an asymmetric bounce is necessary in this scenario.  Assuming the Hubble parameter at $t_e$ to be of the order of $H_e \sim 10^{-55} t_{\rm Pl}^{-1}$ ---as would be the case for a symmetric bounce--- then for the amplitude of the scalar perturbations to match the observed value of $\De_\mR^2 \sim 10^{-9}$, it would be necessary to have a very small value of the sound speed of cold dark matter of $c_s = \ep \sim 10^{-15}$ (assuming $\rho_c$ to be of approximately the same order of magnitude as $\rho_{\rm Pl}$); such a small value of the sound speed parameter typically leads to primordial perturbations with over-large non-Gaussianities.  In order to have a larger (although still small) value of $\ep$ and thus sufficiently small non-Gaussianities, it is again necessary to have an asymmetric bounce.  This likely requires even more asymmetry than is needed to satisfy the condition \eqref{asymmetry}.

Finally, many of the modes observed today reentered the sound Hubble radius during radiation-domination.  However, in order for them to be scale-invariant, in the contracting branch of the universe, they must have exited the sound Hubble radius during matter-domination.  This is yet another reason that an asymmetric bounce is necessary in the $\Lambda$CDM bounce scenario.

\subsection{Numerical Analysis of the Perturbations}
\label{ss.scal-n}

To complete the analysis of cosmological perturbations, in this subsection we perform a numerical computation of the evolution of the primordial curvature perturbations and gravitational waves in the model under consideration. To be consistent with the background numerics, we consider the universe filled with a dust matter field with its energy density evolving as \eqref{rho_tot_CDM}, of which the initial value is the same as the one given in \eqref{para_bg}. Furthermore, we impose the initial conditions of the cosmological perturbations to be vacuum fluctuations during the matter dominated contracting phase (in units where $\hbar=1$),
\be \label{para_pert}
 {\cal R}_k^{ini} \rightarrow \frac{e^{-i \ep k\eta}}{\sqrt{2 \ep k} \, z} ~,~~
 h_k^{ini} \rightarrow \frac{e^{-ik\eta}}{\sqrt{2k} \, a}.
\ee
In addition, we take $\ep = 0.08$ and fix $k = 2.2\times 10^{-7}$ as an example in the detailed calculation. Our numerical results are presented in Fig.~\ref{fig3}.

In the left panel of Fig.~\ref{fig3}, one can see how primordial cosmological perturbations evolve from the sub-Hubble scale to the super-Hubble region in the contracting phase. For a fixed comoving wave number $k$, the curvature perturbation exits the Hubble radius much earlier than the gravitational wave since its oscillation gets squeezed when $k \sim {\cal H} / c_s$ at the sound horizon which is much smaller than the Hubble radius. Consequently, one expects that there ought to be more oscillations in the power spectrum of primordial gravitational waves than that of primordial curvature perturbations. This expectation is exactly verified in the middle panel of the figure, which displays the evolutions of primordial power spectra of both scalar and tensor perturbations.

In the middle part of Fig.~\ref{fig3}, one can see that $\Delta_{\cal R}^2$ experiences only one oscillation and then becomes squeezed very soon with its amplitude increasing until the nonsingular bounce takes place. However, $\Delta_{h}^2$ experiences several oscillations during the contracting phase and only becomes squeezed when the universe is very near the bouncing phase. They both become conserved at super-Hubble scales after the bounce, which can be read from the right regime of Fig.~\ref{fig3b}. Interestingly, one can observe that the magnitudes of the two spectra are comparable (though the amplitude of the scalar mode is slightly larger) during the contracting phase but the amplitude of the scalar spectrum becomes significantly larger than that of tensor spectrum after the bounce, and hence the tensor-to-scalar ratio is suppressed to a small value that is consistent with observations.

This can also be seen in the right panel of Fig.~\ref{fig3} which depicts the dynamics of the tensor-to-scalar ratio $r$ throughout the cosmological bouncing evolution.  In Figs.~\ref{fig3b} and~\ref{fig3c}, we see that while the scalar perturbation mode passes through the bouncing phase in a relatively smooth fashion, the magnitude of the gravitational wave mode is significantly damped.  While the amplitude of the gravitational wave increases somewhat after the bounce, it remains significantly lower than before the bouncing phase.

In the specific example with $\epsilon = 0.08$ being considered, we read approximately $r \simeq 0.016$ from the numerical computation. As we will analyze in the next section on the tensor-to-scalar ratio, this is roughly in the same order of the analytical estimate ($r \simeq 0.012$) given by \eqref{r_ratio}, though the detailed values are not in exact agreement with each other.  We will comment further on this slight numerical discrepancy at the end of this section.

Another important point is that in the numerical computation we have chosen the critical density $\rho_c$ to be very small ($\rho_c\sim10^{-9}$) so that the bounce scale is of order $O(10^{-6})$ Planck mass (this choice for $\rho_c$ significantly lowers the computational cost of the numerics as this causes the bounce to occur at a lower curvature scale).  As a consequence, the amplitude of the power spectrum of primordial curvature perturbations is about $O(10^{-12})$ which is approximately three orders lower than the magnitude of the observed CMB spectrum. The numerical computation performed in this subsection is sufficient to demonstrate the formation of primordial power spectra in the model under consideration and to verify their relations obtained in semi-analytic analyses. However, one can improve the agreement of the amplitude of the spectrum with observations by fine-tuning the values of the parameters in this model.

It is also important to keep in mind that it is possible to obtain a similar amplitude of scalar perturbations even if $\rho_c \sim \rho_{\rm Pl}$, although such a choice would require a smaller value of $|H_e|$ and/or a larger value of $\ep$.

Finally, note that since there are two matter components used in this step of the numerical computation, the sound speed parameter depends on the background evolution approximately as
\be
c_s =
\begin{cases}
\ep & {\rm for~} ~ \Omega_m \simeq 1 \\
\sqrt \omega &{\rm otherwise},
\end{cases}
\ee
where $\ep$ is the small constant sound speed for CDM and $\omega$ is the time-dependent equation of state $\omega = P_{tot} / \rho_{tot}$, which includes the contributions coming from the radiation field and therefore equals $1/\sqrt{3}$ near the bounce when radiation dominates the dynamics of the universe. This time-dependence effect has been taken into account in the above numerical computation.

Finally, these numerical solutions also validate the approximations made in the analytical section.  First, we see that in the near-bounce region it is justified to assume that the modes of interest are in the long-wavelength limit.  Furthermore, we have also numerically checked the difference between the solution obtained in the previous section under the approximation of a discontinuous change of the equation of state, and the result plotted in Fig.\ \ref{fig3b} where no such approximation is made.  We do not include the graph here, as the two curves lie practically one on top of the other, showing the validity of this approximation.  A careful comparison of the two curves shows that the theoretical calculation given in Sec.\ \ref{ss.hom-a} very slightly overestimates the amplitude of $\mR$.

This last point raises an important issue: the results presented in this paper rely either on analytic calculations based on several approximations that are well-motivated but certainly introduce some small errors, or on numerical simulations that, while accurate enough for our purposes here, are not high-precision numerical studies.  Due to the use of these approximations and numerical simulations, the predictions presented in this paper necessarily contain some small errors.  That being said, these small errors are not expected to affect the reliability of the estimates obtained for the predicted observables in the $\Lambda$CDM bounce scenario we consider here, which we estimate (by comparisons between the analytical and numerical results) to be accurate up to an overall factor of approximately 2.

\section{Tensor Perturbations}
\label{s.tens}

It is possible to also calculate the spectrum of tensor perturbations after the bounce, again assuming that the initial state was the quantum vacuum.

The LQC effective Mukhanov-Sasaki equation for tensor perturbations is%
\footnote{It is not possible to use the simplest separate universe models to handle tensor modes as they necessarily require off-diagonal elements in the metric, which do not appear in isotropic space-times.  While it is likely that the separate universe approach could be appropriately generalized by using the anisotropic Bianchi cosmologies to model each `separate universe', this has not been done yet.  Instead, the equation of motion here is derived by demanding that the constraint algebra in the effective theory be anomaly-free.  This procedure, when used to study scalar perturbations, gives the same effective Mukhanov-Sasaki equation as what was obtained in lattice LQC \cite{Cailleteau:2011kr}.}
\cite{Cailleteau:2012fy}
\be
\mu'' - \left( 1 - \f{2 \rho}{\rho_c} \right) \nabla^2 \mu - \f{z_T''}{z_T} \mu = 0,
\ee
where $\mu = h / z_T$, with $h$ being the usual tensor perturbation modes, and
\be
z_T = \f{a}{\sqrt{1 - 2 \rho / \rho_c}}.
\ee

While the `time-dependent potentials' $z''/z$ for scalar perturbations and $z_T''/z_T$ for tensor perturbations are not the same, the most important difference for our purposes is the fact that the sound speed for tensor perturbations is always 1, while the sound speed for scalar perturbations is significantly smaller than 1 during the matter-dominated phase.  This effect strongly suppresses the tensor-to-scalar ratio, as we show below.

With the tensor power spectrum defined as
\be
\Delta_h^2 = \f{k^3}{2 \pi^2} 64 \pi G |h|^2,
\ee
and the tensor-to-scalar ratio $r$
\be
r = \f{\Delta_h^2}{\Delta_\mR^2},
\ee
it is possible to calculate the spectrum of the tensor perturbations in a fashion analogous to Sec.~\ref{s.scal}.  While the calculation is a little long, it is not particularly illuminating as it follows exactly the same steps as the one for scalar perturbations.  While there are a few numerical factors that are different, the procedure is identical and therefore here we will simply state the results.

The power spectrum of the tensor perturbations is found to be almost scale-invariant, although the tilt depends on the value of the effective equation of state at the time that the mode exits the Hubble radius.  Recall that the scalar modes reach the long wavelength limit well before the tensor modes since, during the cold-dark-matter-dominated era, the sound speed for scalar modes is $c_s^{(S)} = \ep$ while $c_s^{(T)} = 1$ for the tensor modes.  Therefore, while the effective equation of state when the scalar perturbations reach the long wavelength limit is $\om_{\rm eff} = -\de$, the effective equation of state when the tensor modes exit the Hubble radius will have changed and will be larger (as the cosmological constant contributes less to the effective equation of state for a smaller scale factor), though still close to zero.  We denote this value of the effective equation of state by $-\de_T$, and we can bound $\de_T$ above by $\de_T \le \de$.  Then the departure from scale-invariance is given by
\be
n_T = - 12 \, \de_T.
\ee
Note that $\de_T$, while expected to be small, may be negative and this would give a slight blue tilt to the spectrum of the tensor perturbations.  Therefore, while near-scale-invariant tensor perturbations are predicted by this model, the exact departure from scale-invariance for the tensor modes will depend on how the effective equation of state varies in time, and this will determine whether there is a small blue tilt or a small red tilt.

Finally, for the particular scenario studied here, the amplitude of primordial gravitational waves is predicted to be very strongly suppressed, with a tensor-to-scalar ratio of
\be\label{r_ratio}
r = 24 \, \ep^3,
\ee
where $\ep$ refers to the sound speed of cold dark matter and $r$ is therefore predicted to be very small.  Note that the tensor-to-scalar ratio is suppressed both by a contribution due to the sound speed of cold dark matter, and also by a further factor of $1/4$ during the bounce due to quantum gravity effects.

There do not appear to be many estimates of the sound speed of cold dark matter in the literature; one interesting reference is \cite{Balbi:2007mz} which provides a bound of approximately $\ep^2 \lesssim 0.03$ (note however that in that paper the authors study a considerably different model from this one).  It is likely that in the future a better estimate for $\ep$ can be found, but nonetheless the constraint $\ep^2 \lesssim 0.03$ already implies an upper bound of $r \lesssim 0.12$ on the tensor-to-scalar ratio, a result which is in agreement with the latest observations \cite{Ade:2013zuv, Hinshaw:2012aka}.

\section{Discussion}
\label{s.disc}

In this paper we have seen how in a contracting universe cosmological perturbations, assumed to be initially in their quantum vacuum state, become scale-invariant if their sound wavelength becomes larger than the Hubble radius when the dynamics of the universe is dominated by cold dark matter.  A small red tilt is generated when the effective equation of state is negative due to the presence of a positive cosmological constant, and a small tensor-to-scalar ratio is predicted.

The scale-invariant perturbations can provide appropriate initial conditions for an expanding universe in order to seed structure formation if there is a bounce to connect the contracting branch of the universe to our current expanding branch.  In the realization of the matter bounce scenario studied here, the bounce occurs due to non-perturbative quantum gravity effects, as captured by LQC, that resolve the classical singularity and provide a quantum bridge between the pre-big-bang and post-big-bang epochs.  Further, the only matter fields present in this model are assumed to be cold dark matter and radiation, together with a positive cosmological constant.

The main predictions depend on four parameters: the scalar index $n_s$ is determined by the effective equation of state at the time when the Fourier modes $v_k$ exit the sound Hubble radius, while the amplitude of the scalar perturbations depend on a combination of the sound speed of cold dark matter $\ep$, the amount of asymmetry in the universe, which can be parametrized by $H_e$ the Hubble rate at matter-radiation equality in the contracting branch, and the matter energy density at the bounce $\rho_c$.  In particular, a symmetric bounce is ruled out and, in order to match observations, it is necessary for matter-radiation equality to occur at higher space-time curvature scales in the contracting branch than in the expanding branch.  This type of asymmetry can be caused by particle production during the bounce, a process which may be important in LQC \cite{Mithani:2014jva}.

There exist several other realizations of the matter bounce scenario, some where the bounce is caused by matter fields that violate energy conditions \cite{Cai:2008qw, Lin:2010pf, Cai:2012va, Cai:2013kja, Cai:2014bea}, and others where it is quantum gravity effects that provide the bounce \cite{Biswas:2005qr, Brandenberger:2009yt, WilsonEwing:2012pu, Brandenberger:2013zea, Peter:2006hx, Odintsov:2014gea}.  There are also some matter bounce scenarios where the gravitational sector is modified not only in the high-curvature regime, but also in the pre-bounce era in order to obtain a matter-like contracting phase without requiring the presence of any matter fields \cite{Leon:2013bra, Bamba:2014mya}.  Of course, there is no \emph{a priori} reason to prefer one realization of the matter bounce scenario over another (other than perhaps the simplicity or the elegance of a particular model).  Instead, it is necessary to determine how these realizations differ in their predictions which can then be compared to observations.

For this reason, it is important to point out that the predictions of the different matter bounce realizations vary in some significant aspects, concerning both CMB experiments \cite{Cai:2008qb, Cai:2009hc, Liu:2010fm, Xia:2014tda, Cai:2014xxa} as well as dark matter searches \cite{Li:2014era, Cheung:2014nxi, Cheung:2014pea}. The recent review \cite{Cai:2014xxa} explains the different predictions of many of the various realizations of the matter bounce scenario and in order to complement that paper, here we shall briefly explain how the predictions of the model studied in this paper differ from the other realizations of the matter bounce scenario.

First, the $\Lambda$CDM bounce model studied in this paper gives a slight red tilt in a natural fashion, something which is absent in many other realizations of the matter bounce (see for example \cite{WilsonEwing:2012pu, Cai:2014bea, Elizalde:2014uba} for specific realizations).  And second, the predicted tensor-to-scalar ratio is very small.  Most realizations of the matter bounce typically predict relatively large tensor-to-scalar ratios, and it is often necessary to assume the presence of a large number of fields (and thus of entropy perturbations as well) in order to decrease the relative amplitude of the tensor perturbations.  This is not necessary here since the tensor-to-scalar ratio is naturally predicted to be small due to the small sound speed of cold dark matter.  Note that this effect of a small $c_s$ strongly damping the value of $r$ has also previously been noticed in a study of the matter bounce scenario in a Bohm-de Broglie quantum cosmology model \cite{Bessada:2012kw}.

Another important point is that this scenario predicts a positive running of the scalar index, $\dd n_s / \dd k > 0$.  This result shows that it is possible for observations to differentiate between the $\La$CDM bounce scenario and inflationary models, where the running of the scalar index is predicted to be negative.  The amplitude of this effect in the $\La$CDM bounce scenario depends on the time evolution of effective equation of state during matter domination $-\de$, which in turn depends quite sensitively on the parameters $\Omega_\La$, $\Omega_m$ and $\Omega_r$ set as initial conditions for the background in the contracting branch.  We leave a detailed study of the relation between these parameters and the amplitude of $\dd n_s / \dd k$ for future work.

There do remain two other important open questions regarding this model that are also left for future work.  The first one concerns the importance of particle production effects during the bounce.  Will there be enough particle production to cause sufficient asymmetry around the bounce point for this model to be viable?  The second, more difficult, open problem is to determine how the presence of anisotropies would modify the predictions of this model.  The Friedmann equation for the mean scale factor in Bianchi models shows that anisotropies dominate the dynamics in the high curvature limit, and we should expect them to typically become important during (and possibly for some time before and after) the bounce in LQC.  Indeed, anisotropies can modify the predictions of a number of cosmological scenarios including inflation \cite{Pitrou:2008gk}.  Despite their importance, anisotropies are often neglected as cosmological perturbation theory becomes considerably more complex in their presence \cite{Den:1986ac, Noh:1987vk, Dunsby:1993fg, Pereira:2007yy};  in particular, for non-vanishing anisotropies the equations of motion for the scalar, vector and tensor perturbations no longer decouple.

Note that if there is an ekpyrotic phase in the contracting branch of the space-time, it is possible to avoid the growth of anisotropies generated during matter contraction \cite{Cai:2012va, Cai:2013kja, Cai:2013vm, Cai:2014zga, Alexander:2014uaa} and then there is no need to include them in the analysis.  However, it is certainly possible (and outside of ekpyrotic models it appears more natural) for anisotropies to grow and dominate the dynamics near the bounce.  In this case, while the anisotropies will be diluted soon after the bounce (see, e.g., Sec.~IIIC of \cite{Gupt:2012vi}), they may change the spectrum of the cosmological perturbations as they evolve through the anisotropy-dominated bounce, and some of these modifications may ultimately be observable today.  For this reason, it is important to allow for anisotropies by generalizing the results obtained here for the case where the background is an anisotropic Bianchi I space-time (rather than flat FLRW), and determining precisely how anisotropies may affect the predictions of this model.

Finally, we conclude with some comments regarding the universality and robustness of the predictions of the $\Lambda$CDM bounce scenario with respect to the physics of the bounce.  In this paper we assumed that the bounce is caused by the quantum gravity effects captured by loop quantum cosmology.  As mentioned above, it is also possible to generate a bounce via other mechanisms, either with matter fields that violate energy conditions, or by modifications to the gravitational action as occur for example in $f(R)$ theories.  An important property of the $\La$CDM bounce scenario is that most of the salient characteristics of the predicted spectra of scalar and tensor perturbations are due to the pre-bounce physics, where quantum gravity and other high energy effects are negligible and therefore the majority of the predictions ---including the small red tilt for the modes that exit the sound Hubble radius when the effective equation of state is slightly negative, as well as the positive running of the scalar index--- are independent of the high-curvature dynamics of the background space-time at the bounce point.  However, there is one effect that is partially due to LQC: the tensor-to-scalar ratio is suppressed by a factor of 1/4 during the bounce by LQC effects (specifically, due to the modifications that arise in the LQC effective Mukhanov-Sasaki equation for tensor perturbations).  Thus, while many predictions of the $\La$CDM bounce are quite robust and are mostly independent of the physics of the bounce, one exception is the predicted value of the tensor-to-scalar ratio which in fact is affected by LQC effects and, if the sound speed of cold dark matter is known, this effect could provide a potentially important observational test for LQC.

\acknowledgments

We would like to thank Robert Brandenberger, Jim Cline, Parampreet Singh, Jun-Qing Xia and Wei Xue for helpful discussions. The work of YFC is supported in part by an NSERC Discovery grant and by funds from the Canada Research Chair program. This work is supported in part by a grant from the John Templeton Foundation.




\raggedright

\end{document}